\documentclass[12pt,preprint]{aastex}

\newcommand{\e}{\epsilon}

\newcommand{\ebw}{e_{Bw}}

\begin{document}

\title{Supranova Model for the Delayed Reddened Optical Excesses Detected in Several GRBs }

\author{Charles D. Dermer}
\affil{E. O. Hulburt Center for Space Research, Code 7653,\\
Naval Research Laboratory, Washington, DC 20375-5352}
\email{dermer@gamma.nrl.navy.mil}

\begin{abstract}

 Excess reddened optical emission has been measured in the power-law
 decay afterglow light curves of several GRBs some tens of days after
 the GRB events.  This emission is claimed to be a signature of a
 supernova (SN) taking place at the same time as the GRB event, in
 support of the collapsar scenario.  Observations of features in
 prompt and afterglow X-ray spectra of GRBs require high density, high
 metallicity material near GRB sources, supporting the supranova (SA)
 scenario.  These conflicting lines of evidence are resolved in likely
 regimes of parameter space if the supernova remnant (SNR) shell is
 illuminated and heated by a pulsar wind (PW), and cools following the
 GRB event. The heating of the SNR shell by the PW is parameterized,
 and simplified expressions for the synchrotron, synchrotron
 self-Compton, and thermal radiation fields within the SNR shell and
 pulsar wind bubble are derived for monoenergetic injection of wind
 electrons.  A cooling SNR shell could produce the excess emission
 detected from GRB 970228, GRB 980326, and GRB 011121. X-ray
 synchrotron radiation from the PW electrons provides an important
 source of ionizing radiation.

\end{abstract}

\keywords{gamma-rays: bursts --- gamma-rays: theory --- radiation 
processes: nonthermal --- pulsars}  

\section{Introduction}

Two leading scenarios to explain the origin of gamma-ray bursts are
the collapsar and supranova (SA) models (see \citet{mes02,der02} for
review).  The collapsar model \citep{ww02,woo93} assumes that GRBs
originate from the collapse of a massive star to a black hole. During
the collapse process, a nuclear-density, several Solar-mass accretion
disk forms and accretes at the rate of $\sim 0.1$-$1~ M_\odot$
s$^{-1}$ to drive a baryon-dilute, relativistic outflow through the
surrounding stellar envelope. The duration of the accretion episode
corresponds to the prompt gamma-ray luminous phase, which is commonly
thought to involve internal shocks.  A wide variety of collapsar
models can be envisaged \citep{fwh99}, but their central feature is
the one-step collapse of the core of a massive star to a black hole.

Optical observations of GRB afterglows reveal significant evidence for
enhanced emission in excess of an extrapolation of the optical
power-law decay spectra in a few cases, namely GRB 980326 (Bloom et
al. 1999), GRB 970228 (Reichart 1999; Galama et al. 2000), and GRB
011121 (Bloom et al. 2002; Price et al. 2002). The spectrum of the
excess is reddened at frequencies above the spectral peak, as might be
expected for a thermal emitter. The peak of the excess emission occurs
tens of days after the GRB event (in the local frame), and the SN
event precedes GRB 011121 by $\sim 3$-5 ($\pm 5$) days \citep{blo02},
consistent with a one-step collapsar model.

The supranova model \citep{vs98} of GRBs involves a two-step collapse
process of an evolved massive star to a black hole through the
intermediate formation of a neutron star with mass exceeding several
Solar masses. The neutron star is initially stabilized against
collapse by rotation, but the loss of angular momentum support through
magnetic dipole and gravitational radiation leads to collapse of the
neutron star to a black hole after some weeks to years.  The
accretion-induced collapse of a neutron star in a binary system could
also form a GRB \citep{vs99}.  A two-step collapse process means that
the neutron star is surrounded by a SNR shell of enriched material at
distances of $\sim 10^{15}$-$10^{17}$ cm from the central source.  The
earlier SN could yield $\sim 0.1$-$1M_\odot$ of Fe in the surrounding
vicinity. The discovery of variable Fe absorption during the prompt
emission phase of GRB 990705 \citep{ama00} and X-ray emission features
in the afterglow spectra of GRB 991216 \citep{pir00} has been
interpreted in terms of the SA model \citep{laz01,bfd02,bal02}. Low
significance X-ray absorption features observed in GRB 970508
\citep{pir99}, GRB 970828 \citep{yos01}, and GRB 000214 \citep{ant00}
can also be modeled within the SA scenario, preferably with
photoionization reflection spectra \citep{laz99}. The multiple
features detected in GRB 011211 \citep{ree02} are also suggested to
arise from an excited SNR shell at $\sim 10^{15}$ cm from the central
source.

A PW and pulsar wind bubble (PWB) consisting of a quasi-uniform, low
density, highly magnetized pair-enriched medium within the SNR shell
is formed by a highly magnetized neutron star during the period of
activity preceding its collapse to a black hole \citep{kg02}.  The
interaction of the PW with the shell material will fragment and
accelerate the SNR shell, and the PW emission will be a source of
ambient radiation that can be Comptonized to gamma-ray energies
\citep{igp02,gg02}.

Here we propose that the nonthermal leptonic PW radiation provides a
source of hard ionizing radiation to heat and photoionize material in
the SNR shell, which is described in terms of a partial covering model
as a result of the strong PW/SNR interaction. This radiation heats the
shell, and the escaping thermalized radiation is seen as the delayed
reddened optical excesses in GRB optical afterglow light curves.  The
source of ionizing nonthermal radiation, which is extinguished when
the GRB occurs, would provide a nonthermal precursor to a GRB, though
at low flux. Various other features of this system are outlined.

\section{Pulsar Wind Physics}

A very rapidly rotating, strongly magnetized neutron star is formed in
the first step of the SA scenario. For a neutron star with surface
polar field of $10^{12}B_{12}$ G, a rotation rate of $10^4\Omega_4$
rad s$^{-1}$, and a radius of 15$r_{15}$ km, the radiated power
$L_0\cong 7\times 10^{44}\; B_{12}^2 r_{15}^6 \Omega_4^4\; \; {\rm
ergs~s}^{-1}$.  The quantity $E_{rot} = {1\over 2} j (GM^2/
c^2)\Omega\;\cong\;2\times 10^{53}\;j_{0.5}M_3^2 \Omega_4\; {\rm
ergs}$ $\equiv 10^{53} E_{53}^{rot}$ is the initial rotational energy
of the neutron star, expressed in terms of the neutron-star mass $M =
3 M_3 M_\odot$ and the dimensionless angular momentum $j_{0.5} =
GM^2/2c$, which remains roughly constant during collapse
\citep{vs98}. The spin-down time $t_{sd} = (E_{rot}/ L_0) \cong
2.8\times 10^8 \; j_{0.5}M_3^2 (B_{12}^2 r_{15}^6 \Omega_4^3)^{-1}$ s.

The torque equation, $\dot \Omega = - K\Omega^n$, where the braking
 index $n = \ddot\Omega\Omega/\dot\Omega^2$, can be solved to give
 $\Omega(t) = \Omega_0(1+t/t_d)^{-1/(n-1)}$, provided that $K$ remains
 constant \citep{ato99}. (For the Crab, $n = 2.5$.) We assume that the
 spin-down power is radiated as a PW in the form of particle and
 Poynting flux, so that the wind luminosity $L_w \cong L_{sd}
 =\;(\xi_e+\xi_p+\xi_B)L_{sd} = L_0/(1+t/t_{sd})^k$, where $k =
 (n+1)/(n-1)$, and we have divided the wind power into leptonic
 (``e"), hadronic (``p"), and electromagnetic (``B") fractions
 $\xi_i,$ $i=e,p,B$.

In the SA model, rotational stability is lost due to spindown, and the
 neutron star collapses to a black hole to produce a GRB at time
 $t_{coll}$. Here we assume $t_{coll} \lesssim t_{sd}$, so that for $k
 > 1$ the wind energy $E_w(t) = \eta\;(k-1)^{-1}L_0 t_{sd}\;[1-(1+(t/
 t_{sd})^{1-k}]\rightarrow \eta L_0t $ when $t \lesssim t_{sd}$, where
 the parameter $\eta$ roughly accounts for the escape of wind energy
 from the SNR shell as well as losses of wind energy due to radiation,
 and is assumed here to be constant with time.  The equation for the
 SNR shell dynamics in a spherical approximation for the SNR shell is
 $M_{SNR}\ddot R = 4\pi R^2 p_w(t)$, where $M_{SNR} = m_{SNR}M_\odot$
 is the SNR shell mass, the wind pressure $p_w(t) \simeq
 E_w(t)/3V_{pwb}$, the PWB volume $V_{pwb} =4\pi R^3(t)/3$, and the
 PWB radius $R(t) = \int_0^t dt^\prime \; v(t^\prime )$
 \citep{igp02}. This equation can be solved to give
\begin{equation}
R(t) = v_0 t (1+ \sqrt{{t\over t_{acc}}})\;,\;{\rm where~}\;\;t_{acc} 
\;= {3 M_{SNR}v_0^2\over 4\eta L_0}\;\cong 1.9\times 10^7\; {m_{SNR}\beta_{-1}^2\over \eta B_{12}^2 r_{15}^6 \Omega_4^4}\;\;{\rm s}\;
\label{R(T)}
\end{equation}
\citep{gg02}. It follows that 
$t_{acc}/ t_{sd}\;= E_{ke}^{SNR}/ \eta E_{rot}\simeq 0.1 E_{52}^{SNR}/
\eta E_{53}^{rot}$, where ${1\over 2} M_{SNR} v_0^2 \cong 9\times
10^{51} m_{SNR}\beta_{-1}^2$ $ \equiv 10^{52}E_{52}^{SNR}$ ergs is the
SNR shell kinetic energy, and $v_0=0.1\beta_{-1}c$ is the SNR shell
coasting speed.  If $\eta \cong 1$, then $t_{acc} \ll t_{sd}$ for
standard values, and shell acceleration must be considered. A highly
porous shell has $\eta \ll 1$. This will occur for strong clumping of
the ejecta, which takes place on the Rayleigh-Taylor timescale $t_{\rm
RT} \simeq \sqrt{x/\ddot R}$, where the characteristic clumping size
scale is $x \equiv f R$ \citep{igp02}. Hence $(t_{RT}/ t_{acc})\;=
f^{1/2}\; (t/ t_{acc})^{3/4}$ in the regime $t \lesssim t_{acc}$. This
shows that small-scale ($f \ll R$) clumping will occur when $f \ll
\sqrt {t/t_{acc}}$. Only a detailed hydrodynamic simulation can
characterize the porosity of the shell due to effects of the pulsar
wind, which $\eta$ parameterizes.  Here we assume, in contrast to the
picture of \citet{kg02}, that a large fraction of the wind energy
escapes the SNR shell and $\eta \sim 0.1$, so that shell acceleration
can be neglected. This effect also causes the shell to become
effectively Thomson thin much earlier than estimated on the basis of a
uniform shell approximation, except for high density clouds with small
covering factor.  A general treatment is needed to treat shell
dynamics, including travel-time delays of the pulsar wind and
deceleration of the SNR shell by the surrounding medium.  When
$t_{coll} \lesssim t_{acc},t_{sd}$, as assumed here, $R \cong v_0 t$.

\section{Radiation Field in SNR Cavity}

The strong PW provides a source of nonthermal leptons, hadrons, and
electromagnetic field. Dominant radiation components considered here
are leptonic synchrotron and SSC radiation, and thermal emission from
the inner surface of the SNR shell which is heated by the nonthermal
radiation.

\subsection{Wind Synchrotron Radiation}

The volume-averaged mean magnetic field $B_w$ in the SNR cavity
powered by the PW is obtained by relating the wind magnetic field
energy density $u_B = B_w^2(t)/8\pi = \ebw u_w(t) = 3 \ebw \eta L_0
t/4\pi R^3$ through the magnetic-field parameter $\ebw$, also assumed
to be constant in time. Thus
\begin{equation}
B_w(t) = {{\cal B}\over t}\;,\;{\rm where~}\; {\cal B} = 
\sqrt{6\ebw L_0 \eta\over v_0^3}\cong 4\times 10^8 \sqrt{\ebw \eta}
\;({B_{12} r_{15}^3 \Omega_4^2 \over \beta_{-1}^{3/2}})\;\; {\rm G-s}\;.
\label{Bwt}
\end{equation}
Let $\gamma_w = 10^5 \gamma_5$ represent the typical Lorentz factors
of leptons in the wind zone of the pulsar. This value should remain
roughly constant when $t \lesssim t_{sd}$. A quasi-monoenergetic
proton/ion wind may also be formed, though we only consider leptonic
processes here, and furthermore do not treat nonthermal particle
acceleration at the PW/SNR shell boundary shock.  When synchrotron
losses dominate, the mean Lorentz factor of a distribution of leptons
with random pitch angles evolves in response to a randomly oriented
magnetic field of mean strength $B$ according to the expression $-\dot
\gamma_{syn} = (\sigma_{\rm T} B^2/ 6\pi m_e c)\gamma^2$, giving
\begin{equation}
\gamma_w = [{1\over \gamma} - T_B\;({1\over t_i} - {1\over t})]^{-1}\;\;,
\label{gammaw}
\end{equation}
where $t_i$ is the injection time, $\gamma_i = \gamma_w$ is the
injection Lorentz factor, and
\begin{equation}
T_B \gamma_w = {\sigma_{\rm T}{\cal B}^2\over 6\pi m_e c}\;
\gamma_w\;\simeq\; 2\times 10^{13} \;\ebw\eta\; 
({B_{12}^2 r_{15}^{6} \Omega_4^4\over \beta_{-1}^3})\;\gamma_5\;{\rm s}\;.
\label{TB}
\end{equation}

Writing the nonthermal electron injection function as
\begin{equation}
{dN_e(\gamma_i,t_i)\over dt_i d\gamma_i} 
\;=\; {\eta\zeta_e L_0 \over m_ec^2 \gamma_w}\;
\delta(\gamma_i-\gamma_w)\;,
\label{dNe}
\end{equation}
the electron Lorentz factor distribution $dN_e(\gamma;t)/d\gamma$ can
be solved to give
\begin{equation}
\gamma^2 {dN_e(\gamma;t)\over d\gamma}\;= \; 
\dot{\cal N}_e T_B \gamma_w^2\;({1\over \hat t}+ 
{1\over \hat \gamma} -1 )^{-2}\;.
\label{g2dNdg}
\end{equation}
In this expression, $\dot{\cal N}_e \equiv \eta \xi_e L_0/(m_ec^2
\gamma_w)$, and we introduce the dimensionless quantities $\hat t = t/
T_B\gamma_w$ and $\hat \gamma = \gamma/ \gamma_w$.  Adiabatic losses,
which are significant on time scales $\sim t/3$, are negligible in
comparison with synchrotron losses of wind electrons when $\hat t \ll
1$, and we restrict ourselves to this regime.

The $\nu L_\nu$ synchrotron radiation flux $\nu L_\nu^{syn} \simeq
{1\over 2} u_B c\sigma_{\rm T} \gamma_s^3 N_e(\gamma_s)$, where
$\gamma_s = \sqrt{\e/\e_B }$, $\e = h\nu/m_ec^2$, $\e_B = B/B_{cr}$,
and the critical magnetic field $B_{cr} = 4.41\times 10^{13}$ G. Thus
\begin{equation}
\nu L_\nu^{syn} (\e ) = c\sigma_{\rm T}\;{B_{cr}^2\over 16\pi}
\;(\dot {\cal N}_e T_B \gamma_w^2)\;({u\over \hat t})^{3/2}\; 
\sqrt{\e }\;({1\over \hat t}+
\sqrt{{u\gamma_w^2\over \e \hat t}}- 1)^{-2}\;\;,
\;\; {\rm for~} \e \leq \e_{max}\;,
\label{nuLnusyn}
\end{equation}
where $u \equiv {\cal B}/\gamma_w B_{cr} T_B \cong 4.5\times 10^{-19}
\beta_{-1}^{3/2}/(\sqrt{\eta\ebw }B_{12} r_{15}^3 \Omega_4^2
\gamma_5$), and
\begin{equation}
\e_{max} \cong \e_B \gamma_w^2 = {{\cal B}
\over B_{cr}}\;{\gamma_w^2\over t} 
\simeq {4.4\times 10^{-9}\over \hat t}\;{\beta_{-1}^{3/2}
\gamma_5\over \sqrt{\eta \ebw } B_{12} r_{15}^3 \Omega_4^2}\;.
\label{emax}
\end{equation}

The synchrotron cooling timescale for wind electrons is $t_{syn} =
\gamma_w T_B \hat t^2$.  In the regime $\hat t \lesssim \beta_0$ where
the wind electrons strongly cool, we can approximate the synchrotron
spectrum by
\begin{equation}
\nu L_\nu^{syn} (\e ) \cong {3\over 8} \xi _e L_0
\;({\e\over \e_{max}})^{1/2}\;H(\e ; \e_0,\e_{max} )\;
\;,
\label{nuLnusynapprox}
\end{equation}
where $H(x;a,b) = 1$ for $a\leq x < b$, and $H=0$ otherwise. The value
of $\hat t$ at $t_{sd}$ is
\begin{equation}
{t_{sd}\over T_B \gamma_w } \;\cong \; 1.4\times 10^{-5} 
\; {j_{0.5}M_3^2\beta_{-1}^3 \over 
\ebw \eta B_{12}^6 r_{15}^{18}\Omega_4^{11} \gamma_5}\;.
\label{ratio}
\end{equation}
Strong cooling holds when $\hat t < t_{sd}/T_B\gamma_w$, noting that
time in physical units can be found from equation (\ref{TB}).

\subsection{Wind Synchrotron Self-Compton Radiation}

The relative importance of synchrotron self-Compton (SSC) to
 synchrotron cooling is given by the ratio $\rho = u_{syn}/u_B$ of the
 synchrotron radiation energy density to $u_B = {\cal B}^2/8\pi t^2$
 (see eq.[\ref{Bwt}]). The PW synchrotron radiation energy density
 $u_{syn}\simeq \kappa (\nu L_\nu^{max})/4\pi R^2 c$, where $\nu
 L_\nu^{max}\cong \xi_e L_0 $ and the parameter $\kappa \geq 1$ is a
 reflection factor such that $\kappa = 1$ corresponds to perfect
 absorption or direct escape (small covering factor) of the PW
 synchrotron radiation, and $\kappa \gg 1$ corresponds to a highly
 reflecting SNR shell with large covering factor.  One obtains $\rho =
 0.033 \kappa \xi_e \beta_{-1}/\ebw \eta$. For large porosity with
 $\kappa \cong 1$ and $\eta \approx 0.1$, we see that the SSC
 contribution is small compared to the synchrotron flux when $\xi_e
 \beta_{-1}/\ebw \lesssim 1$. We restrict ourselves to this regime
 where equation (\ref{gammaw}) is valid.

In the $\delta$-function approximation for the energy gained by a
photon upon being scattered by relativistic electrons \citep{dss97},
with scattering restricted to the Thomson regime, the $\nu L_\nu$ SSC
spectrum from the cooling wind electrons is given by
\begin{equation}
\nu L_\nu^{SSC} \simeq {\sigma_{\rm T}\kappa \xi_e L_0
\over 8\pi v_0^2 T_B }\; {\dot{\cal N}_e \over \hat t^2}\;
\sqrt{{\e\over \e_{max}}}\;\bigl [{1\over a_0(a_0+x)} + a_0^{-2}
 \ln({x\over a_0+x})]|_{x_0}^{x_1},
\label{nuLnuSSC}
\end{equation}
where $a_0 = \hat t^{-1} -1$, $x_0 = \gamma_w \sqrt{\e_0/\e }$, and
$x_1 = \gamma_w \sqrt{\max (\e, \e^{-1}, \e_{max})/\e }$, where we
have used approximation (\ref{nuLnusynapprox}) for the synchrotron
spectrum.

\subsection{Thermal Emission from the Interior of the Shell}

The SNR shell is pictured as having been shredded by the PW and
therefore highly porous, though permeated with small scale ($\sim
10^{12}$-$10^{14}$ cm) density irregularities consisting largely of
metals \citep{dm99,bfd02}.  Note that $0.1 m_{56,-1} M_\odot$ of
$^{56}$Ni provides a radioactive decay power of $\sim 2\times 10^{42}
m_{56,-1}$ ergs s$^{-1}$ before decaying on an $\approx 120$ day
timescale.  The partial covering factor $p_c ~(\leq 1)$ corresponds to
the covering fraction by optically-thick, dense shell
inhomogeneities. Recognizing the severe limitations in the following
expression, we determine the effective temperature $T_{eff}$ in the
interior of the shell through the relation
\begin{equation}
4\pi R^2 \sigma_{\rm SB} T_{eff}^4p_c \cong 
\;{\cal A}p_c\;{\xi_e L_0\over \kappa} + 
2\times 10^{42} m_{56,-1}\exp(-t/120{\rm ~d})\;{\rm ergs~s}^{-1}\;,
\label{Teff}
\end{equation}
where ${\cal A}(\leq 1)$ is an absorption coefficient, dependent in
general on the evolving radiation spectrum, though ${\cal A}$ is here
assumed constant.  The wind heating is more important than heating
from the decay of radioactive Ni when
$({\cal A}/ 0.1) (p_c/ 0.1)\;(\xi_e/ 1/3) \gtrsim$ $ m_{56,-1}
/ B_{12}^2 r_{15}^6 \Omega_4^4$. In this regime,  
\begin{equation}
T_{eff} = ({{\cal A}\xi_e L_0\over 4\pi R^2 \sigma_{\rm
SB}})^{1/4}\approx {825 {\cal A}^{1/4}\over
\sqrt{t/t_{sd}}}\;({\xi_e\over 1/3})^{1/4}\; {B_{12}^{3/2}
r_{15}^{9/2} \Omega_4^{5/3}\over \sqrt{j_{0.5}\beta_{-1}}M_3}\;\;{\rm
K}\;,
\label{Teff1}
\end{equation}
letting $\kappa \approx 1$ because $p_c \sim 0.1$.  For standard
parameters, the shell is heated to $\sim 10^4$ K temperatures when
$t/t_{sd} \ll 1$.  When $B_{12} \gtrsim 4$, this occurs at
$t/t_{sd}\lesssim 1$. The PW heat source is extinguished once the GRB
occurs, and the thermal emission from the SNR shell will decay on the
dynamical timescale $t_{dyn} = R(t)/ c \cong 324
\beta_{-1}\;(j_{0.5}M_e^2/ B_{12}^2 r_{15}^6 \Omega_4^3)\;(t/ t_{sd})$
d, noting that the cooling timescale of a hot shell is generally
shorter than $t_{dyn}$.

\section{Results and Discussion}

Fig.\ 1 shows the total $\nu L_\nu$ flux composed
of synchrotron, SSC, and thermal components, calculated according to
the above simple analysis.  Standard parameters are $j_{0.5}=B_{12} =
r_{15} = \Omega_4 = \beta_{-1} = \kappa=M_3 = m_{52,-1} = \gamma_5=1$,
$\xi_e = e_{Bw} = 1/3$, and $p_c = \eta={\cal A} = 0.1$. The heavy
solid and dashed curves correspond to standard parameters with
$t/t_{sd} = 0.1$ and 0.01 as labeled, with $t_{dyn} = 32.4$ and $ 3.2$
days and $T_{eff} = 1700$ and $5420$ K, respectively ($t_{sd} = 3240$
d). This system is in accord with the standard SA model and might
explain the delayed excess emission in GRB 011121, contrary to the
conclusions of \citet{blo02}.  The set of three dotted curves employ
standard parameters, except that $B_{12} = 4$, and $t/t_{sd} = 1,
0.1,$ and 0.01 as labeled ($t_{sd} = 202$ d), with $t_{dyn} = 20.2,
2.0$, and $ 0.2$ days ($= 17$ ks), and $T_{eff} = 3760, 11900,$ and
$37630$ K, respectively. 

The intense PW X-ray synchrotron emission
provides a source of ionizing photons throughout the SNR shell that
would decay on the timescale $t_{dyn}$ (the line flux would decay on
$\approx 2 t_{dyn}$). The $B_{12} =4, t/t_{sd} = 0.01$ case might
correspond to a GRB 011211-type system. The PW radiation field
provides an additional source of ionizing photons in the SA model
not considered by, e.g., \citet{kn02}, and would be energetically
important for this case if $\gamma_5 \cong 0.1$. A varying ionization
flux impinging on an expanding SNR shell will produce a characteristic
kinematic signature of shell illumination. Our results differ from the
conclusions of \citet{gg02}, because we assume that the SNR shell
highly porous, as in the scenario of \citet{igp02}.

The thin solid curve is the result for standard parameters except that
$B_{12} = 2$, $\gamma_w = 10^6$, and $t = 0.1 t_{sd} = 81$ d.  For
this case, $t_{dyn}=8.1$ d and $T_{eff} = 4360$ K. It is apparent that
a sizeable parameter regime could make quasi-thermal emission between
$\sim 10^{14}$-$10^{15}$ Hz at the level of $\sim 10^{42}$-$10^{43}$
ergs s$^{-1}$ some tens of days after the GRB event.

Thus, even if the GRB takes place months after the SN, a delayed
reddened excess from the cooling shell may be seen, for plausible
parameters, some tens of days after the GRB.  The PW heating is in
addition to the the heating due to the Ni$\rightarrow$ Co
$\rightarrow$ Fe chain, and may relax the need to have $m_{56,-1} \gg
1$, as required to explain the unusually bright light curve of SN
1998bw \citep{iwa98}.  The thermal emission from the SNR shell heated
by the PW is proposed to be the origin of the reddened optical excess
emission detected in GRBs 970228, 980326, and 011121.

This model predicts decaying nonthermal ionizing hard X-ray emission,
which could be detected with INTEGRAL or Swift from nearby $z \lesssim
0.1$-0.3 GRBs when the PW synchrotron flux is at the level $\gtrsim
10^{45}$ ergs s$^{-1}$. A nonthermal optical emission signature would
accompany the reddened excess, with a relative intensity that depends
sensitively on $\gamma_w$. A hard X-ray sky survey instrument such as
EXIST could monitor for the PW nonthermal emission preceding aligned
and off-axis GRBs. The X-ray synchrotron radiation provides a source
of external radiation to enhance photomeson production by energetic
hadrons over the standard GRB model \citep{der02,ad01}.  High-energy
neutrino and GLAST gamma-ray observations can distinguish the
collapsar model and different versions of the SA model (A.\ M.\ Atoyan
\& C.\ D.\ Dermer, 2002, in preparation).

\acknowledgements
{I would like to thank M. B\"ottcher, A.\ M.\ Atoyan, and S.\ D.\ Wick
for discussions and comments. This work is supported by the Office of Naval
Research. }

\clearpage
\begin{figure}
\epsscale{1.0}
\vskip-1.8in
\plotone{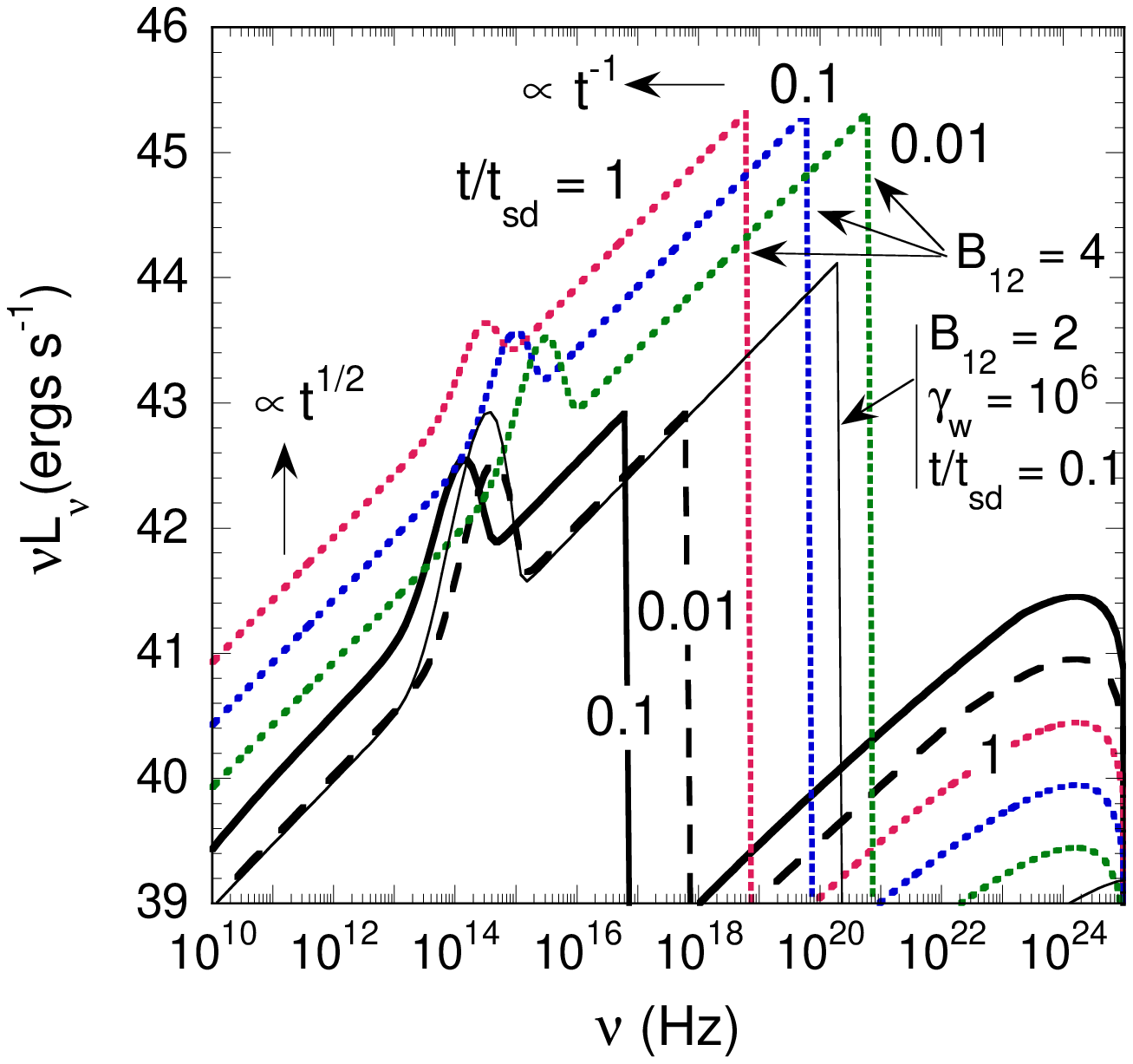}
\vskip-2.0in
\caption{Total model spectral energy distribution 
composed of synchrotron, SSC, and shell thermal
emission from a PW and wind-heated SN shell. Standard parameters,
given in the text, are used, except where noted in the legends.  }
\label{f1}
\end{figure}

%!***********************************************************


\newpage
\begin{thebibliography}{}
\bibitem[Amati et al.(2000)]{ama00} Amati, L.~et al.\ 2000, 
Science, 290, 953 
\bibitem[Antonelli et al.(2000)]{ant00} Antonelli, L.~A.~et 
al.\ 2000, \apjl, 545, L39 
\bibitem[e.g., Atoyan(1999)]{ato99} Atoyan, A.\ M.\ 1999, A\&A, 346, L49
\bibitem[Atoyan \& Dermer(2001)]{ad01} Atoyan, A.\ M., and Dermer, C.\ D. 2001, \prl, 87, 221102
\bibitem[Ballantyne et al.(2002)]{bal02}Ballantyne, D.\ R., Ramirez-Ruiz, E., Lazzati, D., \& Piro, L. 2002, A\&A, 389, L74
\bibitem[Bloom et al.(1999)]{blo99} Bloom, J.~S.~et al.\ 
1999, \nat, 401, 453 
\bibitem[Bloom et al.(2002)]{blo02} Bloom, J.~S.~et al.\ 
2002, \apjl, 572, L45 
\bibitem[B{\" o}ttcher et al.(2002)]{bfd02} B{\" 
o}ttcher, M., Fryer, C.~L., \& Dermer, C.~D.\ 2002, \apj, 567, 441 
%\bibitem[Dermer \& Schlickeiser(2002)]{ds02}Dermer, C.\ D., \& Schlickeiser, R. 2002, \apj, 575, 667
\bibitem[Dermer \& Mitman(1999)]{dm99} Dermer, C.\ D., \& Mitman, K.\ E.\ 1999, Astrophys.\ J., 513, L5
\bibitem[Dermer, Sturner, and Schlickeiser(1997)]{dss97} Dermer, C. D.,  
Sturner, S. J., and Schlickeiser, R. 1993, \apjs, 109, 103
\bibitem[Dermer(2002)]{der02} Dermer, C.\ D., 2002, in Proc.\ 27th ICRC, Hamburg, Germany (7-15 August 2001), astro-ph/0202254
\bibitem[Dermer(2002a)]{der02a} Dermer, C.\ D., 2002, \apj, 574, 65
\bibitem[Fryer, Woosley, \& Hartmann(1999)]{fwh99} Fryer, 
C.~L., Woosley, S.~E., \& Hartmann, D.~H.\ 1999, \apj, 526, 152 
\bibitem[Galama et al.(2000)]{gal00} Galama, T.~J.~et al.\ 
2000, \apj, 536, 185 
\bibitem[Guetta \& Granot(2002)]{gg02}Guetta, D., \& Granot, J. 2002, \mnras, submitted (astro-ph/0208156)
\bibitem[Inoue et al.(2002)]{igp02} Inoue, S., Guetta, D., \& Pacini, F. 2002, \apj, in press
(astro-ph/0111591)
\bibitem[Iwamoto et al.(1998)]{iwa98} Iwamoto, K.~et al.\ 
1998, \nat, 395, 672
\bibitem[K{\" o}nigl \& Granot(2002)]{kg02} K{\" o}nigl, 
A.~\& Granot, J.\ 2002, \apj, 574, 134 
\bibitem[Kumar \& Narayan(2002)]{kn02}Kumar, P., \& Narayan, R. 2002, \apj, in press (astro-ph/0205488)
\bibitem[Lazzati et al.(1999)]{laz99} Lazzati, D., Campana, S., \&
Ghisellini, G.\ 1999, 
\mnras, 304, L31 
\bibitem[Lazzati et al.(2001)]{laz01} Lazzati, D., 
Ghisellini, G., Amati, L., Frontera, F., Vietri, M., \& Stella, L.\ 2001, 
\apj, 556, 471 
\bibitem[M{\' e}sz{\' a}ros(2002)]{mes02} M{\' e}sz{\' a}ros, 
P.\ 2002, \araa, 40, 137 
\bibitem[Piro et al.(1999)]{pir99} Piro, L.~et al.\ 1999, 
\apjl, 514, L73 
\bibitem[Piro et al.(2000)]{pir00} Piro, L.~et al.\ 2000, 
Science, 290, 955 
\bibitem[Price et al.(2002)]{pri02} Price, P.~A.~et al.\ 
2002, \apjl, 571, L121 
\bibitem[Reeves et al.(2002)]{ree02} Reeves, J.~N.~et al.\ 
2002, \nat, 416, 512 
\bibitem[Reichart(1999)]{rei99} Reichart, D.~E.\ 1999, \apjl, 
521, L111 
\bibitem[Reichart(2001)]{rei01} Reichart, D.~E.\ 2001, \apj, 
554, 643 
%\bibitem[Vietri, Perola, Piro, \& Stella(1999)]{vie99} Vietri, M., Perola, C., Piro, L., \& Stella, L.\ 1999, \mnras, 308, L29 
\bibitem[Vietri \& Stella(1998)]{vs98} Vietri, M.~\& Stella, 
L.\ 1998, \apjl, 507, L45
\bibitem[Vietri \& Stella(1999)]{vs99} Vietri, M.~\& Stella, 
L.\ 1999, \apjl, 527, L43 
\bibitem[Yoshida et al.(2001)]{yos01} Yoshida, A.~et al.\ 
2001, \apjl, 557, L27 
\bibitem[e.g.,~Wang \& Woosley(2002)]{ww02}Wang, W., \& Woosley, S.\ E.. 2002, Proc. of 3D Stellar Evolution Workshop, Livermore, CA (July 2002) (astro-ph/0209482)
\bibitem[Woosley(1993)]{woo93} Woosley, S.~E.\ 1993, \apj, 405, 273 
\end{thebibliography}
\end{document}